\documentclass[twocolumn]{aastex63}
\newcommand{\unit}[1]{\, {\rm #1}}
\received{June 3, 2020}
\submitjournal{ApJ}

\shorttitle{SN~2017eaw}
\shortauthors{Weil et al.}

\begin{document}

\title{Detection of Late-Time Circumstellar Interaction of SN~2017eaw in NGC~6946}

\correspondingauthor{Kathryn E.\ Weil}
\email{Kathryn.E.Weil.gr@dartmouth.edu}

\author[0000-0002-8360-0831]{Kathryn E.\ Weil}
\affil{Department of Physics and Astronomy, 6127 Wilder Laboratory, Dartmouth College, Hanover, NH 03755, USA}
\affil{Smithsonian Astrophysical Observatory, 60 Garden Street, Cambridge, MA 02138, USA}

\author[0000-0003-3829-2056]{Robert A.\ Fesen}
\affil{Department of Physics and Astronomy, 6127 Wilder Laboratory, Dartmouth College, Hanover, NH 03755, USA}

\author[0000-0002-7507-8115]{Daniel J.\ Patnaude}
\affil{Smithsonian Astrophysical Observatory, 60 Garden Street, Cambridge, MA 02138, USA}

\author[0000-0002-0763-3885]{Dan Milisavljevic}
\affil{Department of Physics and Astronomy, Purdue University, 525 Northwestern Avenue, West Lafayette, IN 47907, USA}

\begin{abstract}
SN~2017eaw, the tenth supernova observed in NGC~6946, was a normal Type II-P supernova with an estimated 11 -- 13$\unit{M_\sun}$ red supergiant progenitor. Here we present nebular phase spectra of SN~2017eaw at +545 and +900 days post-max, extending approximately 50 -- 400 days past the epochs of previously published spectra. While the +545 day spectra is similar to spectra taken between days +400 and +493, the +900 day spectrum shows dramatic changes both in spectral features and emission line profiles. The H$\alpha$ emission is flat-topped and box-like with sharp blue and red profile velocities of $\simeq-8000$ and $+7500\unit{km}\unit{s}^{-1}$. These late-time spectral changes indicate strong circumstellar interaction with a mass-loss shell, expelled $\sim$ 1700 years before explosion. SN~2017eaw's +900 day spectrum is similar to those seen for SN~2004et and SN~2013ej observed 2 -- 3 years after explosion. We discuss the importance of late-time monitoring of bright SNe~II-P and the nature of pre-supernova mass-loss events for SN~II-P evolution.

\end{abstract}


\section{Introduction} \label{sec:17eaw_intro}
Supernovae (SNe) play an important role for chemical enrichment of the interstellar medium and in their host galaxies evolution including the creation of neutron stars and stellar-mass black holes. Core-collapse SNe arise from the death of massive stars ($\ge8\unit{M_\sun}$) and those that have hydrogen in their spectrum are classified as Type~II events \citep{Fili1997,Arcavi2017,GalYam2017}. 

Further sub-classifications are based on light curve evolution. Historically, Type II-plateau SNe (SNe II-P) have a plateau phase following peak brightness which lasts of order 2 -- 3 months before a linear decline, while Type II-linear SNe (SNe II-L) decline linearly almost immediately after reaching peak brightness. However, recent studies have suggested a more continuous distribution of Type II SNe based on light curve decay rates rather than a bimodal distribution, where the length plateau is related to the mass of the progenitor, with lower mass progenitors having longer plateaus \citep{Anderson2014,Valenti2016,Gutierrez2017}. 

Archival images showing the progenitors of supernovae prior to explosion have led to the identification of red supergiants (RSGs) for the progenitors of SNe~II-P \citep[e.g.][]{VanDyk2003,Smartt2004,MS09,Fraser2012,Fraser2014,Maund2014a,Maund2014b}. The initial masses of these RSG progenitors are typically inferred to be between  $9.5$ -- $16.5\unit{M_\sun}$ \citep[e.g.,][]{Smartt2009,Smartt2015}. This mass range is supported by models of SN~II-P nebular phase spectra \citep{Jerkstrand2012,Jerkstrand2014}. 

As SNe evolve, they become more optically transparent due to decreased ejecta density, which leads to an increasingly emission-line dominated nebular phase starting $\sim150$ -- $200$ days after peak brightness. This nebular phase can provide valuable insights into the kinematic and elemental properties of a SN's ejecta, as well as information about the SN's circumstellar (CSM) environment. Bright supernovae are especially valuable for these studies as they can be observed longer in the nebular phase.

One of the brightest recent SN~II-P was SN~2017eaw in NGC~6946, the 10th supernova in this galaxy in the past century. SN~2017eaw was discovered on 2017 May 14.238 (UT) \citep{Wiggins2017,DS2017} and reached a peak V magnitude of 12.8. Its light curve and spectroscopic evolution followed that of a normal SN~II-P \citep{Cheng2017,Xiang2017,Tomasella2017,Tsvetkov2018,Szalai2019,vanDyk2019,BK2019}.

Archival {\sl Hubble Space Telescope} ({\sl HST}) and Large Binocular Telescope images of NGC~6946 enabled the identification of SN~2017eaw's progenitor as an estimated $11$ -- $13\unit{M_\sun}$ RSG \citep{vanDyk2017,KF2018,Johnson2018,Rui2019}. In addition, archival {\sl Spitzer} imaging also showed that the progenitor was surrounded by a dusty ${\rm T}=960\unit{K}$ shell at $4000\unit{R_\sun}$ \citep{KF2018}. 

Late-time, nebular phase CSM interactions have been reported for a few SNe~II-P. The optical signature of a SN's interaction with surrounding CSM is the appearance of a box-like or horned H$\alpha$ emission feature as the outer layers of hydrogen-rich ejecta are heated by radiation from the forward shock colliding with the CSM \citep{CF03,CF06,CF17,Mili2012}. Box-like or flat-topped emission profiles result from ejecta colliding with shells of CSM, while horned profiles result from collisions with CSM disks \citep{Gerardy2000,Jerkstrand2017}.  The timescale at which the collision occurs allows one to determine the radius of the CSM shell or disk, while the H$\alpha$ velocity width gives information about the CSM density, as denser CSM will decelerate the ejecta more resulting in narrower emission \citep[e.g.,][]{Kotak2009,Andrews2018}. 

Signatures of ejecta-CSM interactions have been observed in a few SNe~II-L/P before $\sim500$ days: SN~2007od \citep{Andrews2010}, SN~2004dj \citep{Andrews2016}, PTF11iqb \citep{Smith2015}, SN~2011ja \citep{Andrews2016}, SN~2013by \citep{Black2017}, and SN~2017gmr \citep{Andrews2019}. However, box-like emission profiles indicating ejecta-CSM interaction with CSM shells have only been observed for three SNe~II-P, and in these cases the ejecta-CSM interaction was observed after $\sim800$ days: SN~2004et \citep{Kotak2009,Maguire2010}, SN~2013ej \citep{Mauerhan2017}, and iPTF14hls \citep{Andrews2018,Sollerman2019}. 

Box-like and horned profiles have been observed more commonly in other types of core-collapse supernovae, for example in Type IIn: SN~1998S \citep{Gerardy2000}, SN~2005ip \citep{Smith2017}, and SN~2013L \citep{Andrews2017}, Type IIb: SN~1993J, \citep{Matheson2000b,Matheson2000a}, SN~2011dh \citep{Shivvers2013}, and SN~2013df \citep{Maeda2015} and Type II-L: SN~1980K \citep{Fesen1990}. SN~1993J, the prototypical SN~IIb, shows boxy H$\alpha$ emission starting at day +669 which lasts until at least day +2500 \citep{Matheson2000b,Matheson2000a}. 

In this paper we present optical spectroscopy of SN~2017eaw at +545 and +900 days after explosion. We found significant changes in spectral evolution between these two epochs indicative of late-time ejecta-CSM interaction. The observations are presented in \S\ref{sec:17eaw_obs} with the results and discussion presented in \S\ref{sec:17eaw_resdis}. Our conclusions are summarized in \S\ref{sec:17eaw_conc}. 

\section{Observations} \label{sec:17eaw_obs}

As part of an optical survey of recent core-collapse SNe in nearby galaxies, we obtained photometric and spectroscopic observations of SN~2017eaw in NGC~6946. Using the Hiltner 2.4m telescope at MDM Observatory equipped with the Ohio State Multi-Object Spectrograph \citep[OSMOS;][]{Martini2011} and an ITL $4064 \times 4064$ CCD, R-band photometry was obtained on 7 November 2018, 14 December 2018 and 3 May 2019. Additional photometry observations using a filter matching the {\sl HST} WFPC2 F675W filter were obtained on 30 September and 28 October 2019. The F675W is a broadband red continuum filter covering the majority of the same emission features as a standard R band filter but with greater sensitivity of the weak, broad  [\ion{Ca}{2}] and [\ion{O}{2}] emission blend. Below we treat F675W images as roughly equivalent ($\pm 0.25$ mag) to images taken in R band. Observations were reduced using the OSMOS\footnote{\url{https://github.com/jrthorstensen/thorosmos}} imaging reduction pipeline in Astropy \citep{AstropyCiteA,AstropyCiteB}. Standard methods of aperture photometry were followed in combination with data from \citet{BK2019} to calibrate our images and measure magnitudes of SN~2017eaw.

\begin{figure*}
\centering
\includegraphics[width=0.95\textwidth]{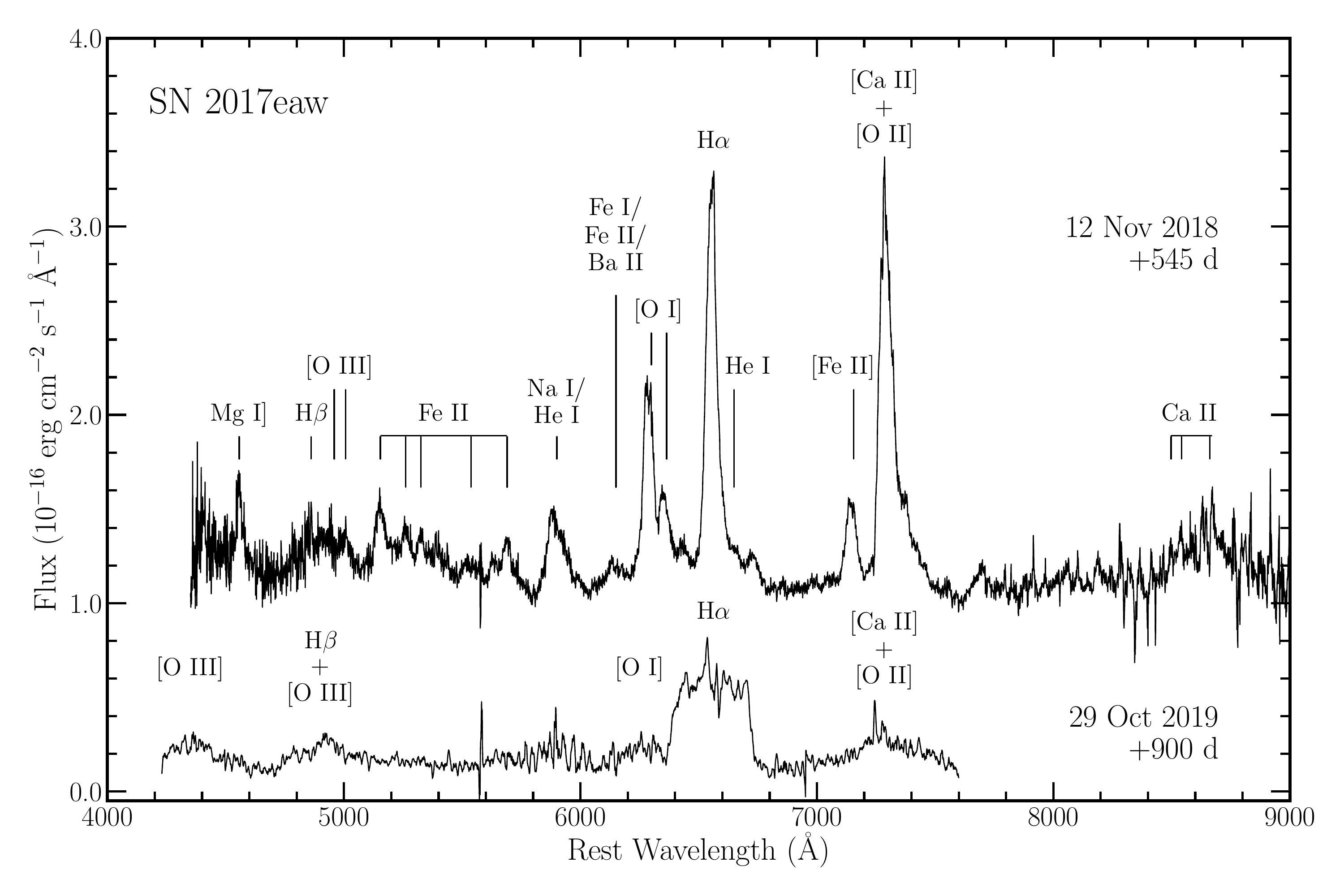}
\caption{Spectra of SN~2017eaw taken at +545 and +900 days. The +545 day spectrum is offset by a constant and smoothed by a boxcar of width 1. The +900 day spectrum is smoothed by a boxcar of width 7. Major emission features for each spectrum are marked. \label{fig:spec_comp}}
\end{figure*}

Optical spectra of SN~2017eaw were taken on 7 -- 9 November 2018 using OSMOS and a 1\arcsec\ wide slit with a red VPH grism. Exposure times were 3000$\unit{s}$ or 4000$\unit{s}$. Observations were reduced using the OSMOS spectral reduction pipeline in Astropy \citep{AstropyCiteA,AstropyCiteB} and calibrated using Hg and Ne lamps and spectroscopic standard stars \citep{Oke74,Massey90}. To improve the signal to noise of the final spectrum, three individual observations were combined to a single 11,000 second exposure. We adopted a time since explosion of +545 days\footnote{We adopted an explosion date of 2017 May 12.2 (JD 2,457,885.7) \citep{vanDyk2019} throughout this paper which is approximately one day earlier than the value used in \citet{Rui2019} and \citet{Szalai2019}.} using the average date of the three individual exposures: 8 November 2018. 

R band images of NGC~6946 taken in early Fall 2019 revealed SN~2017eaw was still visible. Using the MMT 6.5m telescope equipped with Binospec \citep{Fabricant2019}, we obtained a spectrum on 29 October 2019, approximately +900 days after explosion. Observations consisted of $3\times1260\unit{s}$ exposures with a 1\arcsec\ slit and the 270 line grating with $1\farcs23$ seeing and thin cirrus. Data were reduced using the Binospec pipeline \citep{BinospecCite} and calibrated with Ne lamps and night sky lines, and flux calibrated with standard star observations.

\section{Results and Discussion} \label{sec:17eaw_resdis}

SN~2017eaw's evolution followed that of a standard Type II-P supernova before day $\sim+500$ with similar photometric and spectroscopic properties to SN~2004et \citep{Tsvetkov2018,Rho2018,Rui2019,Szalai2019,vanDyk2019,Tinyanont2019,BK2019}. There was little evidence of CSM interaction for SN~2017eaw in all previously published results, which cover SN~2017eaw's evolution up to day +594. A small and brief magnitude increase was observed in the optical light curve of SN~2017eaw between 6 -- 10 days after explosion, similar to that observed in SN~2012aw, which has been attributed to CSM interaction \citep{Szalai2019}. There was also a weak narrow H$\alpha$ emission feature which only appeared in the 1.4 day spectrum \citep{Rui2019}.

The +545 day nebular phase spectrum of SN~2017eaw shown in Figure~\ref{fig:spec_comp} is dominated by ejecta emission lines with strong  H$\alpha$ 6563\,\AA, [\ion{O}{1}] 6300, 6364\,\AA, and blended [\ion{Ca}{2}] 7291, 7323\,\AA\ and [\ion{O}{2}] 7320, 7330\,\AA\ emission. The spectral shape and observed emission lines closely resemble the +482 and +490 day spectra, but the flux has decreased by approximately 50 percent \citep{vanDyk2019,Szalai2019}. Other strong line emissions are marked in Figure~\ref{fig:spec_comp}, which agree with other nebular phase emission spectra for SNe~II-P \citep[e.g.,][]{Jerkstrand2012,Silverman2017,vanDyk2019}.

The [\ion{O}{1}] and blended [\ion{Ca}{2}] + [\ion{O}{2}] emission are asymmetric with blue-shifted emission profiles. The [\ion{O}{1}] 6300\,\AA\ emission peaks on the blue-side at $-600\unit{km}\unit{s}^{-1}$ while on the red-side the peak is at $3000\unit{km}\unit{s}^{-1}$. For the blended [\ion{Ca}{2}] +  [\ion{O}{2}] feature, the emission peaks at $-1600\unit{km}\unit{s}^{-1}$, with a red-component at approximately $2000\unit{km}\unit{s}^{-1}$. Modeling this blended profile as two components, one red and one blue, the blue component contained 60 percent of the total flux. The asymmetries and the shifted emission profiles indicate the presence of dust in the oxygen-rich ejecta layers, in agreement with dusty progenitors suggested by \citet{Khan2017}, \citet{KF2018}, \citet{Szalai2019}, and \citet{vanDyk2017,vanDyk2019}. 

Further evidence for dust in SN~2017eaw's ejecta is indicated by {\sl Spitzer} infrared observations up to +560 days post explosion, from which dust masses of $10^{-4}\unit{M_\odot}$ of silicate dust and $10^{-6}\unit{M_\odot}$ carbonaceous dust were estimated \citep{Tinyanont2019}. This amount of dust is 100 to 10,000 smaller than model predictions for SNe~II-P. However, more dust may form over time  and/or the dust may have formed quickly and currently be obscured in optically thick regions so that the observed dust mass would increase as the ejecta becomes optically thin \citep{Tinyanont2019}.  

\begin{figure}
\centering
\includegraphics[width=0.45\textwidth]{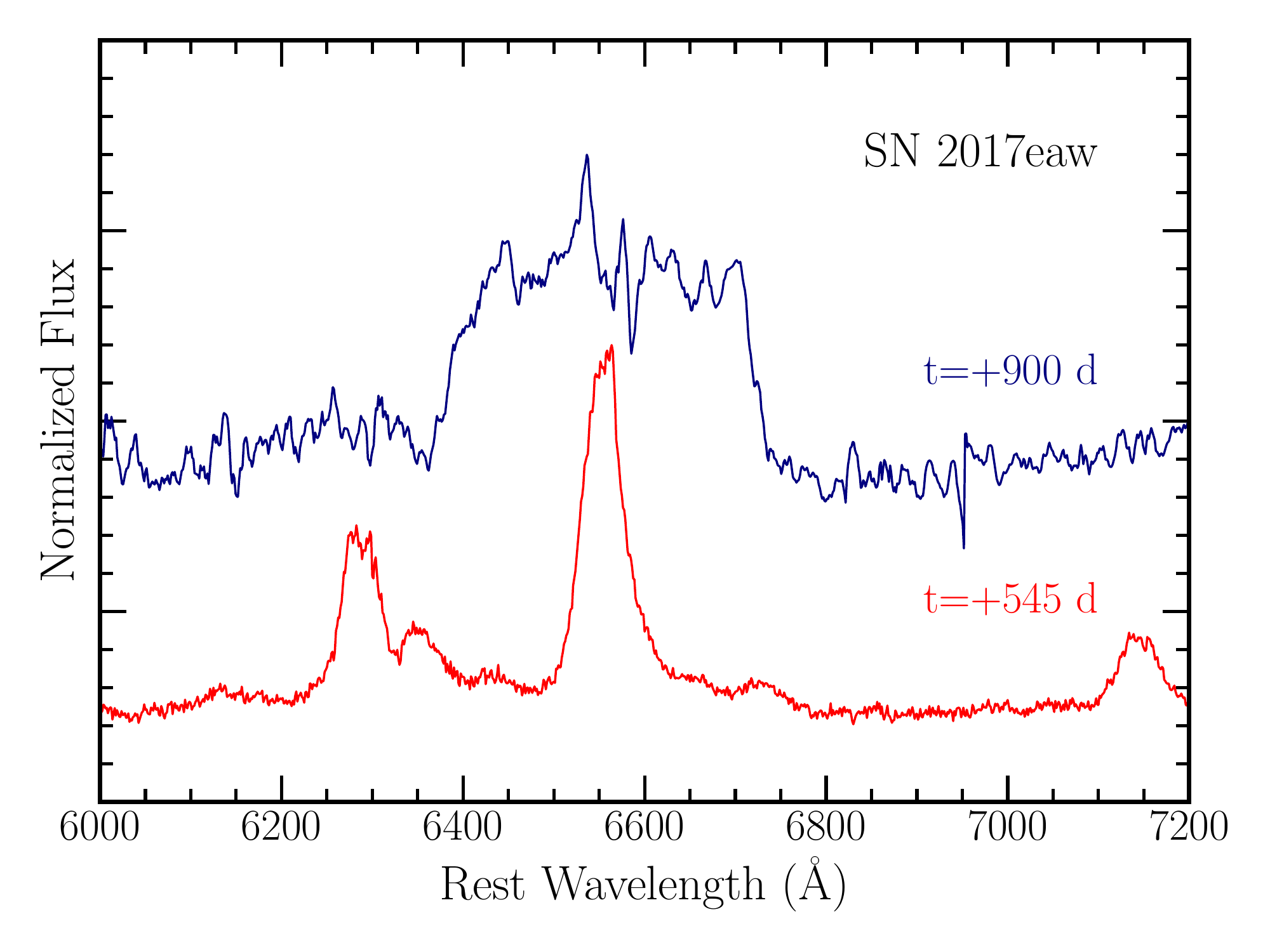}
\caption{Comparison of the H$\alpha$ and [\ion{O}{1}] emission profiles of SN~2017eaw taken at +545 and +900 days. The H$\alpha$ broadens from HWZI~$\sim2000\unit{km}\unit{s}^{-1}$ on day +545 to $\sim8000\unit{km}\unit{s}^{-1}$ on day +900. \label{fig:halpha_spec_evol}}
\end{figure}

\subsection{Ejecta-Circumstellar Interaction} \label{subsec:17eaw_inter}

Our +900 day spectrum shown in Figure~\ref{fig:spec_comp} revealed a drastic change in the supernova's optical spectrum. The H$\alpha$ emission now appears broad and box-like while the [\ion{O}{1}] and [\ion{Ca}{2}] + [\ion{O}{2}] blend have broadened but decreased in relative strength compared to H$\alpha$. In addition, weak, blended, and broad H$\beta$ and [\ion{O}{3}] emission are also present. The narrow emission features from the blends of \ion{Fe}{2} below 6000\,\AA\ are no longer visible and neither is the blend of \ion{Fe}{1}, \ion{Fe}{2}, \ion{Ba}{2} at approximately 6100\,\AA. The significant changes to the H$\alpha$ and [\ion{O}{1}] emission profiles is highlighted in Figure~\ref{fig:halpha_spec_evol}.

The boxy H$\alpha$ profile is strong evidence of ejecta-CSM interaction. The emission extends between approximately $-8000$ and $+7500 \unit{km}\unit{s}^{-1}$. The fact that this box-like feature is so broad (HWZI $\sim8000\unit{km}\unit{s}^{-1}$) suggests that the highest velocity ejecta are running into high density CSM before the ejecta have become optically thin. Further, the flat-top of this feature indicates the CSM is a thick shell rather than a disk or torus which would have horn-like emission profiles that with increased emission at the highest velocities \citep{Gerardy2000,Jerkstrand2017}. 

The larger bumps in the otherwise flat-topped H$\alpha$ profile appear to be real as they are stronger and wider than simply noise on top of the flat emission profile. These features may indicate either clumpy ejecta interacting with a thick shell of CSM material, or smooth ejecta interacting with a clumpy shell of CSM emission \citep{Gerardy2000,Jerkstrand2017}. 

\begin{figure}[t]
\centering
\includegraphics[width=0.47\textwidth]{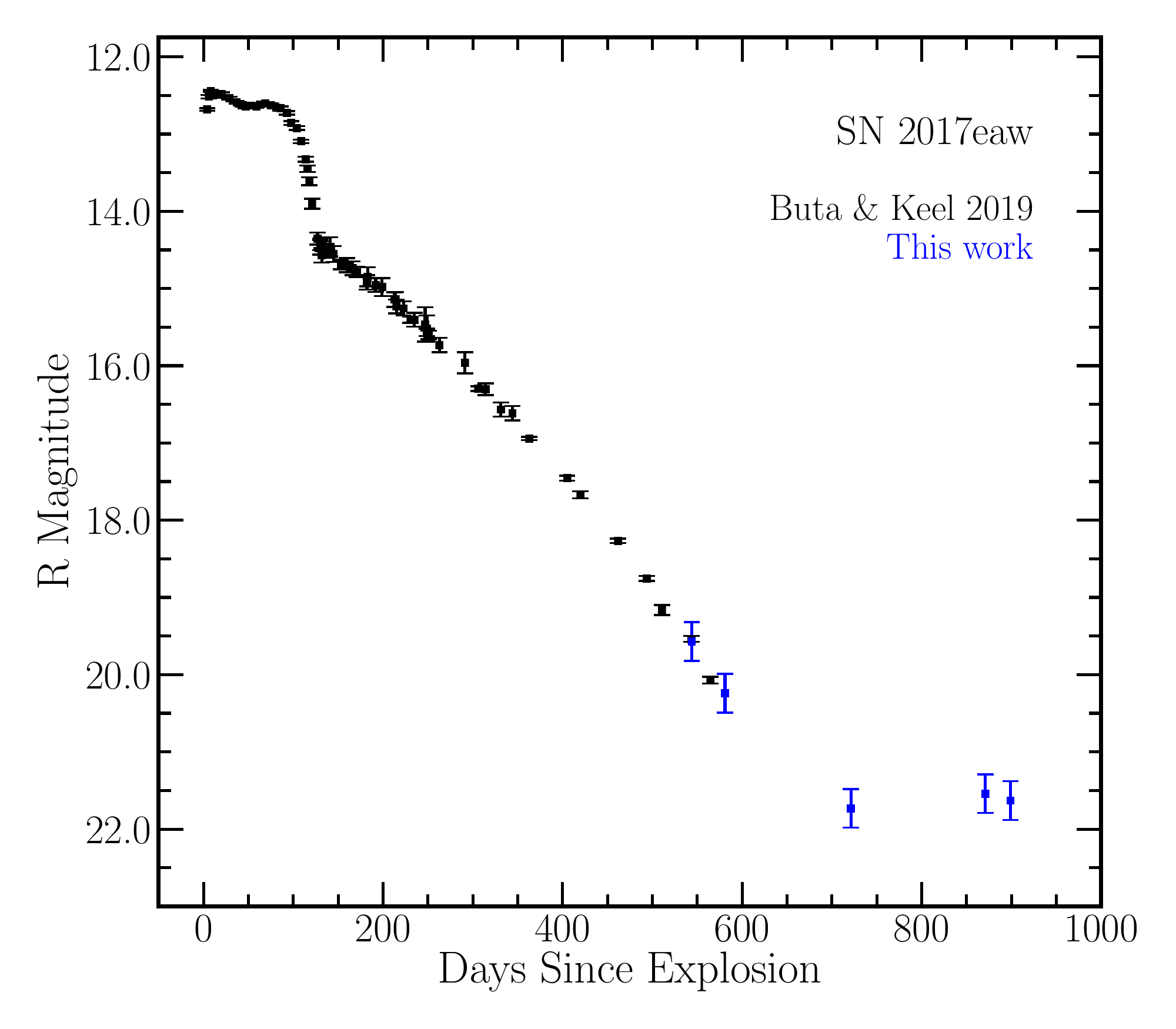}
\caption{R-Band light curve of SN~2017eaw to +899 days since the explosion. This paper's data extends the light curve of \citet{BK2019} shown in black. At day +721, the R-band magnitudes begin to level off possibly signalling the beginning of the ejecta-CSM interaction, however no spectra are available at that time. \label{fig:Rband_LC}}
\end{figure}

We can constrain the approximate epoch for ejecta-CSM interaction via the R-band light curve evolution, as shown in  Figure~\ref{fig:Rband_LC}. The late-time nebular phase R-band evolution between days +290 and +564 followed a decay rate of $1.469\pm0.031\unit{mag}\unit{(100\, d)}^{-1}$ \citep{BK2019}. However, at +721 days, SN~2017eaw had an observed R-band magnitude of $21.73\pm0.25$ indicating a change in the decay rate. Between +720 and 871 days the light curve was flat with a slight decrease $\approx 0.1\unit{mag}$ between +871 and +899 days. No observations were obtained between +590 and +721 days, as SN~2017eaw was behind the Sun. Pre-explosion {\sl HST} images of the SN~2017eaw site indicate that no bright optical sources are located within a radius of $2\arcsec$ \citep{vanDyk2019}, meaning that late-time measurements do not have significant contributions from underlying host emission.  

One can calculate the electron density, $n_e$, of the ionized hydrogen ejecta, from the recombination time $\tau_r=(\alpha_A n_e)^{-1}$, where $\alpha_A$ is the recombination coefficient of hydrogen. Assuming a temperature of $10,000\unit{K}$, a recombination time of 100 days, the electron density is $2.77\times10^{5}\unit{cm}^{-3}$. We choose a recombination time of 100 days, as the light curve data indicates a flattening between day +581 and +721. However, as we have no spectral data between day +545 and +900, the recombination time could be somewhat longer.

\begin{figure}[t]
\centering
\includegraphics[width=0.47\textwidth]{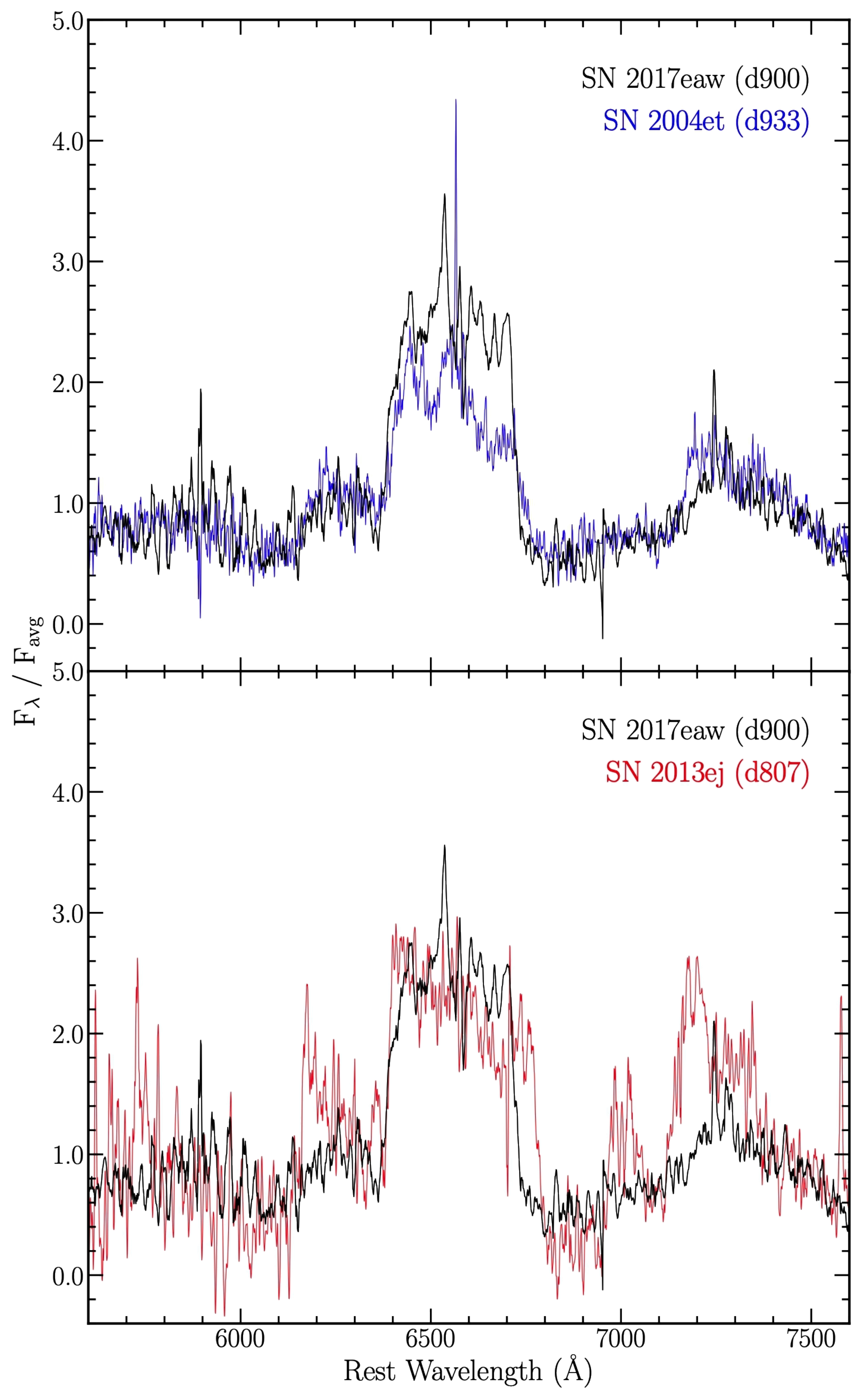}
\caption{900 day spectrum of SN~2017eaw (black) compared to late-time spectra of the other two SN~II-P showing box-like profiles: SN~2004et \citep[top blue;][]{Kotak2009} and SN~2013ej \citep[bottom red;][]{Mauerhan2017}. The spectra have been normalized and a small constant was added to SN~2013ej so that the normalized continuum emission below 6000\,\AA\ was approximately equal to SN~2017eaw. SN~2004et and SN~2013ej spectra were obtained from the WISeREP \citep{Yaron2012}. \label{fig:17eaw_comp}}
\end{figure}

\begin{figure}[t]
\centering
\includegraphics[width=0.47\textwidth]{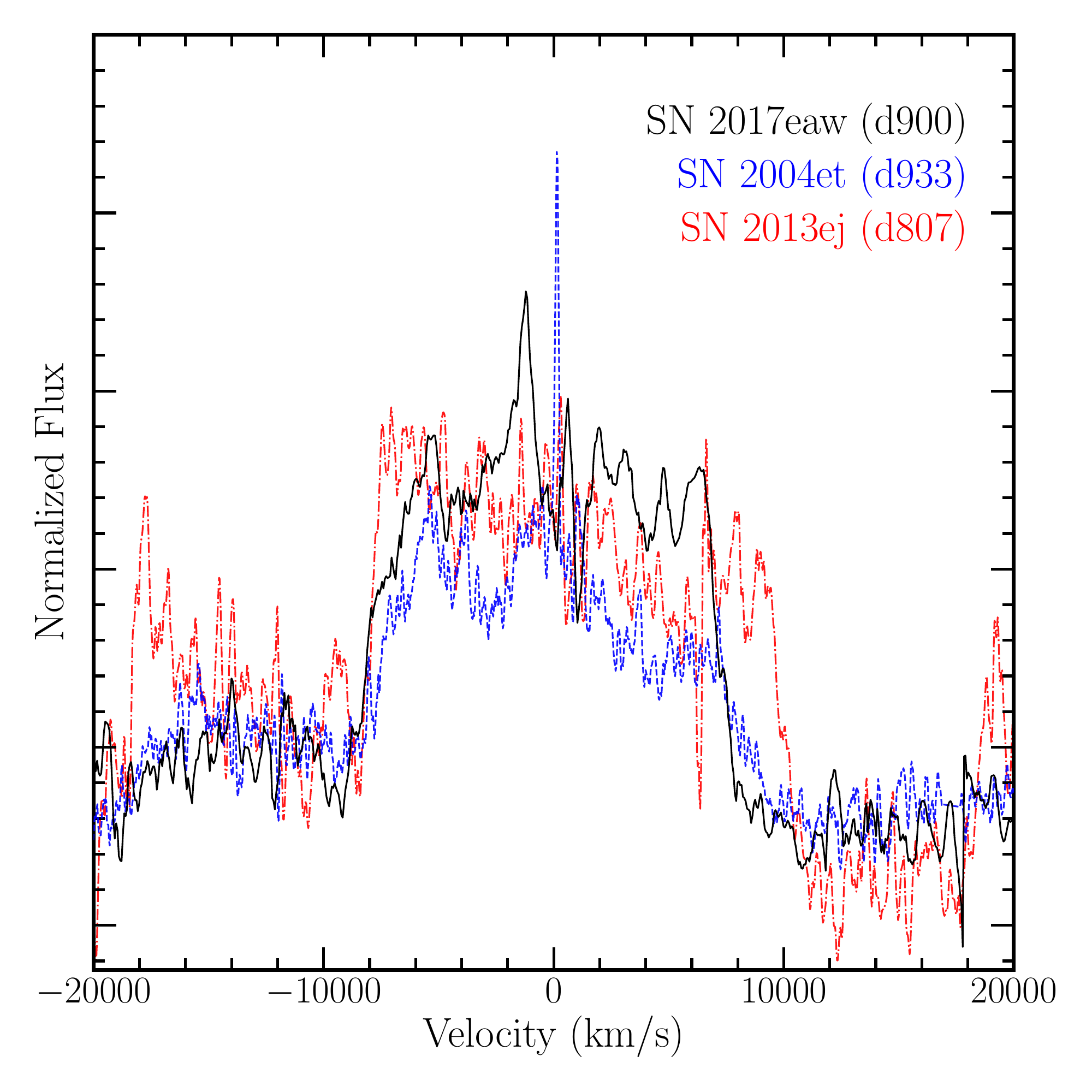}
\caption{H$\alpha$ line emission relative to expansion velocity for SN~2017eaw (black), SN~2004et (blue, dashed) and SN~2013ej (red, dashed-dotted). All three SNe have box-like H$\alpha$ emission profiles which rise at $\sim$ $-8000\unit{km}\unit{s}^{-1}$. SN~2017eaw and SN~2004et emission extends to $\sim$7500$\unit{km}\unit{s}^{-1}$ while SN~2013ej extends further to the red to approximately 10,000$\unit{km}\unit{s}^{-1}$. Note: the blue side of the H$\alpha$ emission feature overlaps with the red side of the [\ion{O}{1}] 6300, 6364\,\AA\ emission. \label{fig:halpha_vel}}
\end{figure}

We can also estimate the mass within the ionized hydrogen using 
\begin{equation}
{\rm M_{H^+}}=\frac{m_p4\pi d^2F^c({\rm H}\alpha)}{h\nu_{{\rm H}\alpha}\alpha_{{\rm H}\alpha}^{eff}(H^0,T_e)N_e}
\end{equation}
where $m_p$ is the mass of the proton, $F^c({\rm H}\alpha)$ is the extinction corrected H$\alpha$ flux, $h\nu_{{\rm H}\alpha}$ is the energy of an H$\alpha$ photon, and $\alpha_{{\rm H}\alpha}^{eff}(H^0,T_e)$ is the effective recombination coefficient of H$\alpha$ \citep{MelnikCopetti2013}. Using the electron density found from recombination, the extinction corrected flux of $2.3\pm0.3\times10^{-14}\unit{erg}\unit{s}^{-1}\unit{cm}^{-2}$, a distance to NGC~6946 of $7.73\pm0.78\unit{Mpc}$ \citep{vanDyk2019}, we estimate $\sim 1.9\unit{M_\sun}$ of ionized hydrogen ejecta. This estimate is in agreement with SNe~II-P/II-L hydrogen mass envelope estimates which range from $1$ -- $10\unit{M_\sun}$ \citep{Capp2001}, while more recent classifications have defined SN~II-P explosions as having hydrogen envelope masses as low as ${\rm M_H}\ge0.3\unit{M_\sun}$ \citep{Limongi2017}. This is a lower limit on the mass of the hydrogen envelope as we assumed the entire hydrogen envelope was fully ionized to make the initially density estimate. If, however, it is only partially ionized, then the mass of the hydrogen envelope would be higher. 

\begin{figure}[t]
\centering
\includegraphics[width=0.47\textwidth]{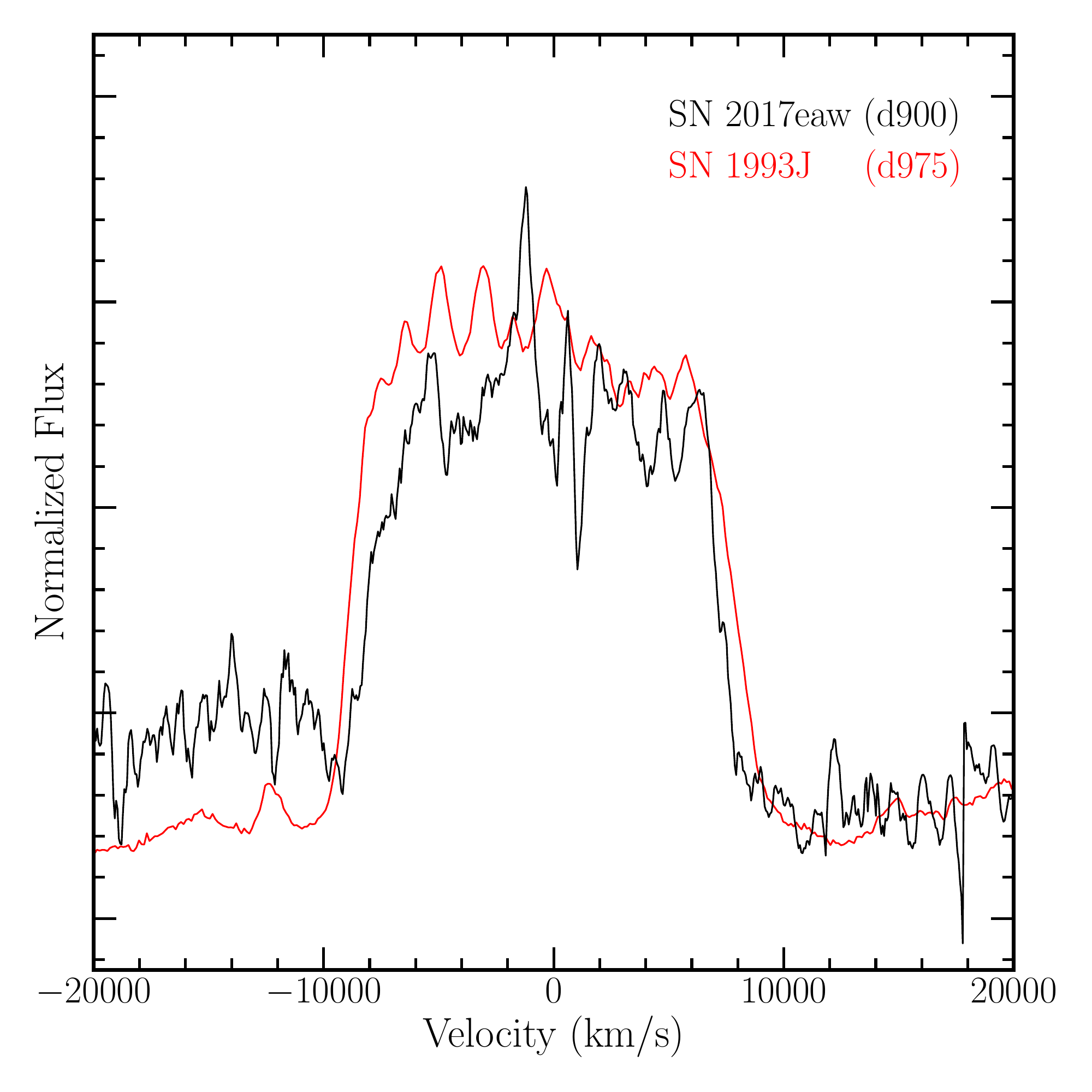}
\caption{H$\alpha$ line emission relative to expansion velocity for SN~2017eaw (black) and the Type IIb, SN~1993J \citep[red,;][]{Matheson2000b,Matheson2000a}. The spectra have been normalized such that continuum emission to the red of H$\alpha$ is approximately equal. Note: the blue side of the H$\alpha$ emission feature overlaps with the red side of the [\ion{O}{1}] 6300, 6364\,\AA\ emission. The SN~1993J spectrum  was obtained from the WISeREP \citep{Yaron2012}. \label{fig:Int_Comp}}
\end{figure}

\subsection{Similarities to SN~2004et and SN~2013ej} \label{subsec:17eaw_04et13ej_comp}

A few other SNe~II-P observed late in the nebular phase have also shown box-like H$\alpha$ profiles indicating ejecta-CSM interactions. These include SN~2004et \citep{Kotak2009,Maguire2010}, SN~2013ej \citep{Mauerhan2017}, and iPTF14hls \citep{Andrews2018,Sollerman2019}. Box-like emission profiles were first observed on day +823 for SN~2004et, on day +807 for SN~2013ej \citep{Mauerhan2017}, and on day +1153 for iPTF14hls \citep{Andrews2018}. 

Figure~\ref{fig:17eaw_comp} shows a comparison SN~2017eaw's spectrum at day +900 with SN~2004et at day +933 (top panel) and SN~2013ej at day +807 (bottom panel), while Figure~\ref{fig:halpha_vel} shows the H$\alpha$ emission feature for these three SNe in terms of radial velocity. The spectra of SN~2017eaw and SN~2004et showed nearly identical broad, box-like H$\alpha$ emission with the same relative strength of [\ion{O}{1}]/H$\alpha$. In addition, the [\ion{O}{1}] emission profiles have very similar widths and similar blueshifted emission profiles. For SN~2013ej, its H$\alpha$ emission on the red side was broader than SN~2017eaw and its [\ion{O}{1}]/H$\alpha$ ratio was stronger than that observed for SN~2017eaw. 
SN~2004et had an H$\alpha$ emission profile with a HWZI of $\sim8500\unit{km}\unit{s}^{-1}$ \citep{Kotak2009}, while SN~2013ej had a HWZI of $\sim9000\unit{km}\unit{s}^{-1}$ \citep{Mauerhan2017} and iPTF14hls, (not shown in these figures) had a HWZI of $\sim1500\unit{km}\unit{s}^{-1}$ \citep{Andrews2018}. 

We note that SN~2017eaw's boxy H$\alpha$ emission also resembles the boxy emission from the Type IIb SN~1993J at day +975 \citep{Matheson2000b,Matheson2000a} as shown in Figure~\ref{fig:Int_Comp}. Compared to SN~2017eaw, SN~1993J's oxygen-rich ejecta emission are weaker compared to its H$\alpha$ emission. For SN~1993J the boxy H$\alpha$ emission was clearly observed as early as day +300 \citep{Finn95}, when [\ion{O}{1}] and H$\alpha$ emission were equal in strength. The [\ion{O}{1}] emission faded more rapidly compared to the H$\alpha$ emission in SN~1993J. 
The H$\alpha$ emission of SN~1993J is slightly wider than that observed for SN~2017eaw, but also relatively flat. We note that for SN~1993J the box-like H$\alpha$ emission feature has been modelled having significant contribution from [\ion{N}{2}] at these late-times due to the low hydrogen envelope mass in SN~IIb events \citep{Jerkstrand2014}. 

We now focus on comparing SN~2017eaw with SN~2004et and SN~2013ej which show comparable expansion velocities and time frames for CSM interaction. Unlike what is observed for both SN~2004et and SN~2013ej, the H$\alpha$ profile of SN~2017eaw appears virtually flat, with no hint of decreased flux on the receding red side compared to those on the near blue facing side \citep[e.g.,][]{Mauerhan2017}. This suggests that there is not a significant amount of dust in the CSM or ejecta which would absorb the emission from the far side of the supernova. Furthermore, SN~2017eaw does not show a multi-peaked H$\alpha$ profile like that observed for SN~2004et and SN~2013ej \citep{Kotak2009,Mauerhan2017}. 

\begin{figure}[t]
\centering
\includegraphics[width=0.47\textwidth]{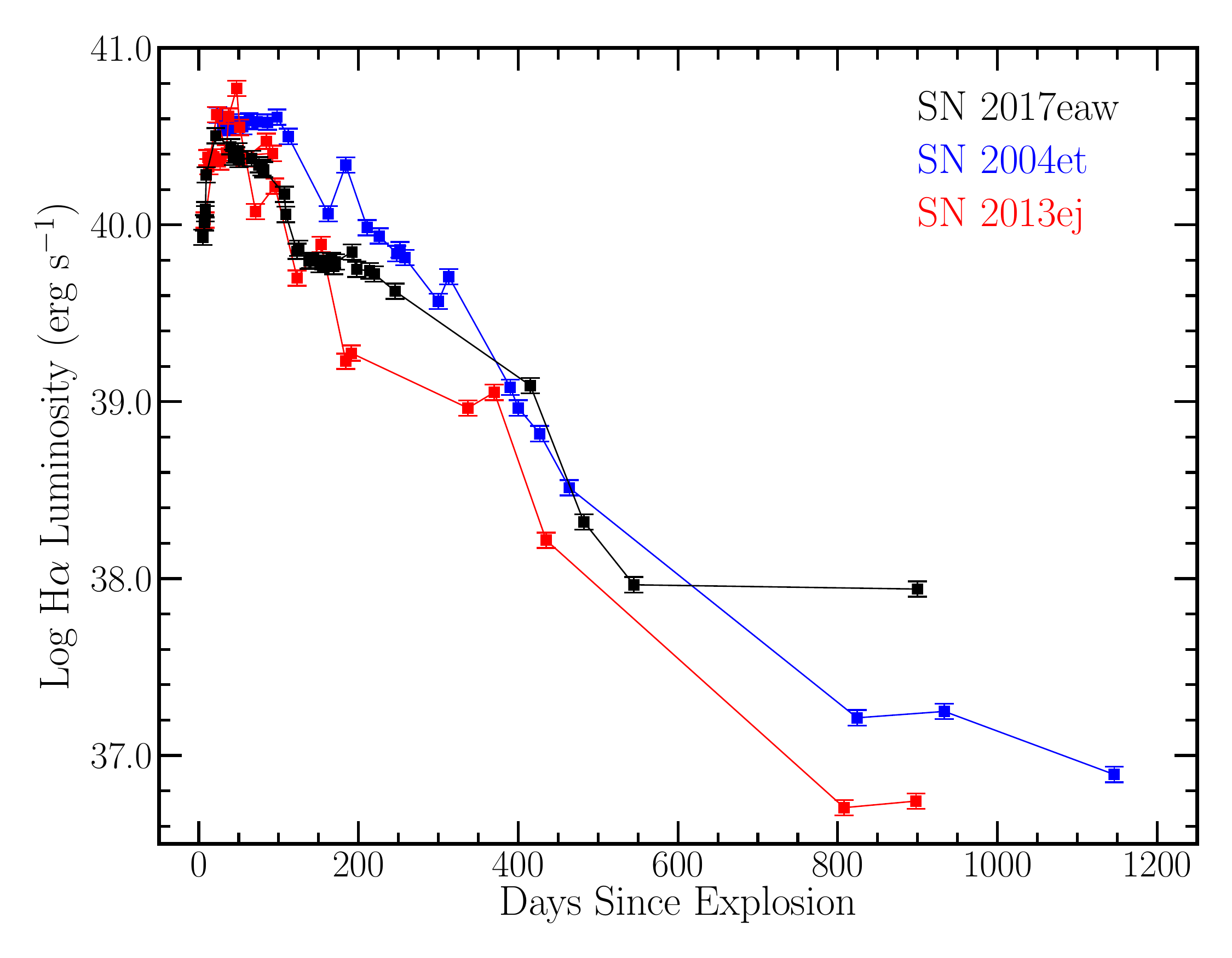}
\caption{Observed H$\alpha$ luminosity of SN~2017eaw (black squares) as a function of age compared with SN~2004et \citep[blue squares;][]{Sahu2006,Kotak2009}, and SN~2013ej \citep[red squares;][]{Valenti2014,Mauerhan2017}. SN~2017eaw luminosity measurements before +481 days were determined from the spectra presented in \citet{vanDyk2019}. Between approximately +481 and +900 days the H$\alpha$ emission remained almost constant. SN~2004et and SN~2013ej H$\alpha$ luminosity measurements were made using spectra obtained from the WISeREP \citep{Yaron2012}. Error bars represent 20 percent error in the flux measurements arising from  uncertainties in the line shape and absolute flux calibration of the individual spectra. \label{fig:halpha_evol}}
\end{figure}

The H$\alpha$ luminosity of SNe~II-P can be used to determine the ejecta nickel mass based on the well-modeled exponential decay rate in the nebular phase \citep[e.g.,][]{Chugai1990}. Following the plateau phase, the H$\alpha$ luminosity of SN~2017eaw, SN~2004et and SN~2013ej exponentially declined until CSM interaction is observed in their spectra, between +800 and +900 days, see Figure~\ref{fig:halpha_evol}. The H$\alpha$ luminosities were measured from previously published spectra of SN~2004et \citep{Sahu2006,Kotak2009} and SN~2013ej \citep{Valenti2014,Mauerhan2017} obtained from the WISeRep database \citep{Yaron2012}. For SN~2004et and SN~2017eaw the luminosity was determined using the \citet{vanDyk2019} $7.73\pm0.78\unit{Mpc}$ distance to NGC~6946. For SN~2017eaw the H$\alpha$ flux at +900 days is about 90 percent of the flux at +545 days, which does not follow the exponential decay predicted and previously observed throughout the rest of the nebular phase.

Infrared (IR) studies of SN~2017eaw between days +100 to +600 showed a similar evolution to SN~2004et which experienced an IR rebrightened at day +1000 shortly after showing CSM interaction in the optical \citep{Tinyanont2019}. SN~2013ej also showed a rebrightening around +800 days, approximately the same time as the broadening and flattening of the H$\alpha$ profile \citep{Mauerhan2017}. The IR brightening in SN~2004et and SN~2013ej cannot be explained solely with IR echos from preexisting dust in the CSM. The more likely scenario is that the MIR increase is from heated dust formed in the CDS as a result of the ejecta-CSM collision \citep{Kotak2009,Mauerhan2017}. Follow-up IR imaging of SN~2017eaw would likely show a similar rebrightening as the ejecta interacts with the CSM. 

\subsection{Characteristics of the Circumstellar Material} \label{subsec:17eaw_csm_char}

We can estimate the radius for the CSM shell surrounding SN~2017eaw by assuming the maximum observed ejecta velocity at day +900 has not greatly changed since the start of CSM interaction. The radius of the CSM shell is then given by $r_{\rm shell}=vt$, where $v$ is the ejecta velocity and $t$ is the date of initial ejecta-CSM interaction.
From R-band photometry, the ejecta-CSM interaction occurred on or before day +721, and, and using the H$\alpha$ emission velocity from our +900 day spectrum as the velocity of ejecta $8500\unit{km}\unit{s}^{-1}$, the radius of the CSM shell is approximately $5.3\times10^{16}\unit{cm}$ or $0.017\unit{pc}$. 
Interestingly, the size of this CSM shell is similar to the $8\times10^{16}\unit{cm}$ shell estimated for SN~2004et \citep{Kotak2009}. Assuming typical red supergiant (RSG) wind speeds of $10\unit{km}\unit{s}^{-1}$, the mass-loss episode responsible for this material would have had to occur $\sim1700\unit{yr}$ before SN~2017eaw exploded. 

Due to long term monitoring of NGC~6946 for SNe or failed supernovae, there exists  archival optical and infrared images of SN~2017eaw's progenitor for nine years prior to explosion. The progenitor showed little variability, ruling out eruptive mass loss episodes for approximately a decade prior to explosion \citep[e.g.,][]{Johnson2018,Tinyanont2019}. The CSM around SN~2017eaw was likely sculpted by photoionized trapped RSG wind with similar to structure to that observed around Betelgeuse, with estimated mass-loss rates  $\dot{M}\sim(0.9-5)\times10^{-6}\unit{M_\odot}\unit{yr}^{-1}$ \citep{KF2018,Tinyanont2019}.

In order to exhibit box-like emission features, the cooling time must be shorter than the adiabatic time scale to enable the formation of the cold dense shell behind the reverse shock \citep{CF03,CF06,CF17,Kotak2009}. From an analysis of SN~2017eaw, like that done for SN~2004et, we estimate the cooling time as $t_c=4.6\times10^{-3}(\dot{M}_{-5}/u_{\rm w1})^{-1}V_{\rm s4}^{5.34}t_{days}^2\unit{days}$, where $\dot{M}_{-5}$ is the mass loss rate in units of $10^{-5}\unit{M_\sun}\unit{yr}^{-1}$, $u_{\rm w1}$ is the wind velocity in units of $10 \unit{km}\unit{s}^{-1}$ and $V_{\rm s4}$ is the ejecta velocity in units of $10^{4}\unit{km}\unit{s}^{-1}$ \citep{CF03,Kotak2009}. Since the box-like emission was present in SN~2017eaw's day +900 spectrum, the shock must still be radiative. Using the width of the H$\alpha$ emission $V_{\rm s}\approx 10^{4}\unit{km}\unit{s}^{-1}$ and the typical RSG wind speeds of $u_{w}=10\unit{km}\unit{s}^{-1}$, we estimate the mass loss rate of the progenitor as $\dot{M}\approx3\times10^{-6}\unit{M_\sun}\unit{yr}^{-1}$ in agreement with previous results derived from X-ray luminosities shortly after outburst and progenitor SED analysis.

\subsection{Frequency of SNe~II-P Showing Late-Time CSM Interactions} 
\label{subsec:17eaw_late_intr}

Analysis of 38 SNe~II-P in the nebular phase up to day +500 did not find any showing box-like emission profiles \citep{Silverman2017}. This raises the question of whether bright Type II-P objects like SN~2004et, SN~2013ej, and SN~2017eaw which have observations past day +500 are unique in showing interaction or if there is simply an underlying observational bias of too few very late-time observations.

Since the majority of SNe~II-P are not observed 2 -- 3 years after explosion, some or maybe most may undergo similar CSM interactions and we simply have not realized it due to a lack of very late-time observations. To answer the question more SNe~II-P events need to be studied in the 2 -- 3 year time frame. We note that all three cases where late-time CSM interactions were observed, the SNe~II-P were relatively bright; SN~2017eaw reached $V = 12.8$ mag, SN~2013ej reached $12.5$ mag, and SN~2004et reached $12.6$ mag. 

Consequently, it is worthwhile to photometrically and spectroscopically follow the next bright SN~II-P event. Moreover, since SN~2004et is still visible \citep{Long2019}, other bright SN~II-P events with similar ages should be looked at again to investigate if they might also still be visible. Although any late-time emission, if present, will be quite faint, such additional late-time detections will give us useful information to better understand the CSM environments around SNe~II-P. 

\section{Conclusions} \label{sec:17eaw_conc}
SN~2017eaw's optical spectrum dramatically evolved between day +545 and +900, developing a box-like, flat-topped H$\alpha$ emission profile, indicative of ejecta-CSM interaction. The H$\alpha$ emission at day +900 dominated the spectrum relative to the oxygen-rich ejecta emission which was strong in earlier phase spectra. SN~2017eaw is just the third SN~II-P with observed ejecta-CSM interaction with box-like emission profiles after +500 days.  However, unlike like the other late-time SNe~II-P spectra, SN~2017eaw showed a fairly flat-topped profile indicating little dust is present in the CSM shell.

Given the similarity of SN~2017eaw's evolution with that of SN~2004et up to day +900, we can predict some of SN~2017eaw's future spectroscopic evolution. Spectra of SN~2004et taken at 10.2 years resembled its day +933 spectrum, in that  broad, box-like H$\alpha$ emission was still observed while oxygen emission increased in relative strength to the H$\alpha$ \citep{Long2019}. It is therefore likely that the broad H$\alpha$ profile observed in the +900 day spectrum of SN~2017eaw will also still be sustained for several years to come. In any case, even if SN~2017eaw's luminosity fades below practical spectroscopic levels, continued photometric monitoring of its evolution would be worthwhile.

\acknowledgments
We wish to thank Eric Galayda and the MDM and MMT staffs for assistance with our observations and John Thorstensen for his assistance in the reduction of some of the data used in this paper. 
K.~E.~W.\ acknowledges support from Dartmouth's Guarini School of Graduate and Advanced Studies, and the Chandra X-ray Center under CXC grant GO7-18050X. 
D.~J.~P.\ acknowledges support from NASA contract NAS8-03060.
D.~M.\ acknowledges support from the National Science Foundation under Award No.\ PHY-1914448.
Some of the observations reported here were obtained at the MMT Observatory, a joint facility of the Smithsonian Institution and the University of Arizona.

\facilities{Hiltner (OSMOS), MMT (Binospec)}

\software{Astropy v4.0 \citep{AstropyCiteA,AstropyCiteB}, Binospec Pipeline v1.0-20190502 \citep{BinospecCite}, OSMOS Pipeline (thorosmos:
\url{https://github.com/jrthorstensen/thorosmos})}

\bibliography{kweil}
\bibliographystyle{aasjournal}

\end{document}